\documentclass[twocolumn, switch]{article}

\usepackage{arxiv}

\usepackage[utf8]{inputenc} 
\usepackage[T1]{fontenc}    
\usepackage{hyperref}       
\usepackage{url}            
\usepackage{booktabs}       
\usepackage{amsfonts}       
\usepackage{nicefrac}       
\usepackage{microtype}      
\usepackage{lipsum}		
\usepackage{graphicx}
\usepackage{natbib}
\usepackage{doi}
\usepackage{caption}
\usepackage{subcaption}
\usepackage{float}
\usepackage{icomma}
\usepackage{amsmath}
\usepackage{tabularx}
\usepackage{array, booktabs, makecell}
\usepackage{siunitx, mhchem}
\usepackage{enumitem}
\usepackage[title]{appendix}
\usepackage{siunitx}

\title{Efficient Stochastic Simulation of Network Topology Effects on the Peak Number of Infections in Epidemic Outbreaks}

\author{ \href{https://orcid.org/0000-0002-6600-2637}{\includegraphics[scale=0.06]{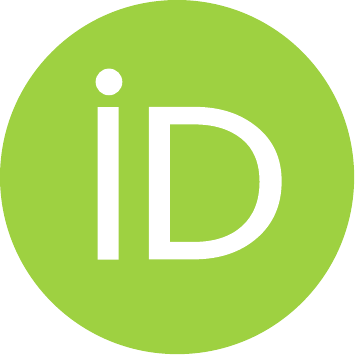}\hspace{1mm}Yulian Kuryliak}\\
	Institute of \\Computer Science and Information Technologies\\
	Lviv Polytechnic National University\\
	12 Stepan Bandera street, 79000 Lviv, Ukraine\\
	\texttt{yulian.kuryliak.kn.2017@lpnu.ua} \\
	\And
	\href{https://orcid.org/0000-0002-7342-2090}{\includegraphics[scale=0.06]{orcid.pdf}\hspace{1mm}Michael Emmerich} \\
	Institute of Advanced Computer Science\\
	Leiden University\\
	Niels Bohrweg 1, 2333CA Leiden, The Netherlands\\
\& Faculty of Information Technology, \\University of Jyvaskyla, \\P.O. Box 35 (Agora), FI-40014
Jyvaskyla, Finland\\
\texttt{m.t.m.emmerich@liacs.leidenuniv.nl} \\
	\And
	\href{https://orcid.org/0000-0003-4040-4467}{\includegraphics[scale=0.06]{orcid.pdf}\hspace{1mm}Dmytro Dosyn} \\
	Institute of Computer Science and Information Technologies\\
	Lviv Polytechnic National University\\
	12 Stepan Bandera street, 79000 Lviv, Ukraine\\
	\texttt{dmytro.h.dosyn@lpnu.ua} \\
}

\renewcommand{\shorttitle}{Network Topology Effects on the Peak Number of Infections in Epidemic Outbreaks}

\begin{document}
\twocolumn[ 
  \begin{@twocolumnfalse} 
  
\maketitle

\begin{abstract}
This paper investigates the effect of the structure of the contact network on the dynamics of the epidemic outbreak. In particular, we focus on the peak number of critically infected nodes (PCIN), determining the maximum workload of intensive healthcare units which should be kept low. As a model and simulation method, we develop a continuous-time Markov chain (CTMC) model and an efficient simulation based on Gillespie's Stochastic Simulation Algorithm (SSA). This methods combines a realistic approximation of the stochastic process not relying on the assumptions of mean field models and large asymptotically large population sizes as in differential equation models, and at the same time an efficient way to simulate networks of moderate size.
The CTMC simulation is implemented in python and integrated in a dashboard that can be used for interactive exploration and it is made openly available.
In our analysis, we focus on the question how the network topology influences the dynamics of the outbreak and the PCIN. Virus propagation is compared on random graph models featuring a selected range of complex network topologies: Erdős–Rényi, Watts-Strogatz, Barabási–Albert and complete graph (Clique). 
Simulations are performed in networks with $200$, $500$, $1000$, $2000$, and $10000$ nodes with the same average degree of a node. Based on this, our aim is to look at interpretable graph features, such as average path length and clustering, to explain how systemically the network topology influences the PCIN.

We study age- and gender-determined and weighted characteristics of nodes on the PCIN as well as the correlation of macroscopic graph characteristics such as the clustering coefficient and the average shortest path length. The analysis uses the data of the demographic distribution of Ukraine as of 2020 and data on mortality from COVID-19 in Ukraine, as of December 16, 2020. 
In networks of moderate size, incorporating the correct demographic characteristics has a small effect on the number of critically infected. 
More importantly, the simulations show that the increase of the average shortest path length is significant on the reduction of the PCIN, whereas other characteristics such as clustering and age distribution, are of lesser importance. 
\end{abstract}
\keywords{
    epidemic outbreak \and
    complex networks \and
    network topology \and
    contact process \and
    peak number of infected nodes \and
    Gillespie's algorithm \and
    Continuous Time Markov chains
} 
\vspace{0.35cm}

  \end{@twocolumnfalse} 
] 

\section{Introduction}
Predicting and managing the dynamics of infectious diseases is a problem of high urgency. The recent outbreak of the COVID-19 pandemic has increased interest in epidemiology, and in particular on the question how the structure of social networks influences the spread of a disease. 

It is important to start the fight against the virus as soon as it is detected, but it usually takes a long time to invent a vaccine, approve it, and distribute it~\cite{flaxman2020estimating}. It is therefore important to contain the spread of the virus until the time when vaccination becomes available. 
The recent COVID-19 pandemic has shown that it is of paramount importance to reduce the number of patients that have to be treated at the same time in intensive care units (ICUs) to avoid the risks that hospitals run out of capacity. To reduce this number, non-pharmaceutical means, such as contact restrictions, will be effective\cite{non-pharmaceutical-interventions}. The question, which contact restrictions are most effective, is a topic of ongoing research~\cite{flaxman2020estimating}.

Classical epidemiology has mainly focused on idealized homogeneous network structures such as complete graphs or networks where each person has about the same number of contacts.
However, a more detailed look at how changes in the contact network topology will affect the spread of an epidemic is necessary. To gain insights into methods for effectively slowing down the spread of the virus in the network, it will be useful to conduct a study into the sensitivity of the rate of spread of the virus to the topology of the network. The existing literature on this topic is mainly focused on asymptotical analysis~\cite{pastor2015epidemic}, i.e. asymptotically large populations, or the early stage of the spread of an epidemic, where the reduction of the largest eigenvalue of the contact network (adjacency matrix) plays a crucial role~\cite{emmerich2020multiple, van2015epidemic}. However, once an epidemic is already spread out across a network, other dynamics need to be taken into account. This topic, however, received relatively small attention in the literature~\cite{pastor2015epidemic} and the results are mainly based on time-discrete models~\cite{Chen2020Time-DependentCOVID-19,ACHTERBERG2020} which may not consider the continuous the important aspect of time in a statistical sound way, or on continuous-time differential equation models that do not take the network structure into account.

Another downside of existing simulation models is that they focus mainly on the epidemic threshold (or reproduction number, which is inversely proportional to it). Although this number certainly is of interest, in a real pandemic other factors deserve more attention when it comes to managing the outbreak:
According to research~\cite{McCabe2021}, forecasting and limiting the peak loads on the hospitals is one of the main tasks in a pandemic. An important indicator that depends on the basic epidemic parameters (topology of the contact network and the rate of infection) is the peak number of simultaneously infected nodes (PCIN). By keeping this number below a critical value it is made sure that hospitals have sufficient capacity to treat all patients with a severe course of the infectious disease and avoid triage situations.

The recent work comprises already a few examples of studies that consided \emph{Continuous Time Markov Chains} for modeling pandemics (but without a focus on network topology influences):
In~\cite{marwa2019ctmc}, a cholera epidemic outbreak was simulated using  (CTMCs) for a SIR model, with nodes of the network that may be susceptible,  infected with symptoms, infected without symptoms, and excluded. In~\cite{romeu2020markov}, CTMCs were used to model the distribution of COVID-19 based on already known statistics. A similar simulation was performed in~\cite{xie2020novel}. Both studies use the SIR model and all three do not take into account network topology and population demographics, but do not consider aspects of network topology.
Other studies emphasize the importance of network topology  , but use simulations that make some simplifying assumptions concerning the simulation method, such as discrete time, mean field assumptions, or asymptotically large population sizes. See \cite{ACHTERBERG2020} for a review of such studies.

The objective of this paper is to provide a realistic, yet efficient, method for simulation of the spreading process, and first results on the effects of network topology, with a focus on the peak number of infected individuals.
Instead of asymptotic analysis (such as differential equations and mean-field models), we propose using the stochastic simulation algorithm (or G`illespie's algorithm~\cite{gillespie2007stochastic}) for simulating CTMCs. This method is realistic, encompasses all stages of an outbreak, flexible and efficient in assessing the effect of network topology on contact networks of moderate to large size. We show that the state-space explosion, that causes the CTMC model based simulations to typically suffer from state-space explosion, can be avoided for the given epidemiological model. In contrast to mean-field methods such as differential equations, the stochastic simulation-based analysis also has the advantage that error margins of the model can easily be assessed, which allows for a robust risk assessment.
In addition, we seek to explain our findings by using graph-features that can be interpreted by non-experts in network science, thereby making the results accessible to a broader range of decision makers. For the same reason, we will make the developed software available as an open-source and public domain dashboard software that makes it possible to interactively explore different scenarios and incorporate new data sets and parameters.

\section{Methodology}
\subsection{Epidemiological Model}
To solve problems related to the analysis of the dynamics of the spread of infectious diseases in a population, experts usually are considering certain generalized models according to which individuals of such a population are in one of three main possible states: 1) susceptible to infection (S), 2) infected and contagious to others (I) and 3) removed from the list of susceptible and infected individuals due to the acquisition of immunity or death due to the sometimes fatal course of the disease (R). According to this division, the main epidemiological models are the SI, SIS, and SIR models~\cite{pastor2015epidemic}.
For diseases with a high rate of spread and fast occurrence of symptoms, models are usually used that do not take into account natural mortality and fertility, as well as population aging, i.\,e., it is assumed that the demographic distribution is stable throughout the epidemic.

Due to the need to model the spread of COVID-19 among the population, the SIR model was selected as the most appropriate. 
To make the epidemiological model closer to COVID-19 we identify 2 infected states: simply infected and critically infected. Simply infected units(IU) - are people who are able to overcome the disease without severe symptoms and people who do not develop symptoms but are contagious. We assume that people are in this state for 10 days. Critically infected (also, intensive care units, ICU) - are people who will die without intensive care. We assume that people are in critically infected state for 14 days. According to the WHO (\url{https://www.who.int/indonesia/news/detail/08-03-2020-knowing-the-risk-for-covid-19}), there are about 20\% of people who need medical care. \footnote{We have used mortality data for different groups distinguished by age and gender. We add to the mortality of the available data 20\% to account for uncertainties in the data acquisition.}


Although COVID-19 has a certain incubation period, it is difficult to determine, therefore, we consider a person contagious immediately after infection. The possibility of returning the removed persons to a state of susceptibility due to the gradual loss of acquired immunity is also not taken into account , as within the duration of time that we simulate, immunity will most likely be preserved among recovered individuals.

\begin{figure}[h]
\centerline{\includegraphics[width=.5\textwidth]{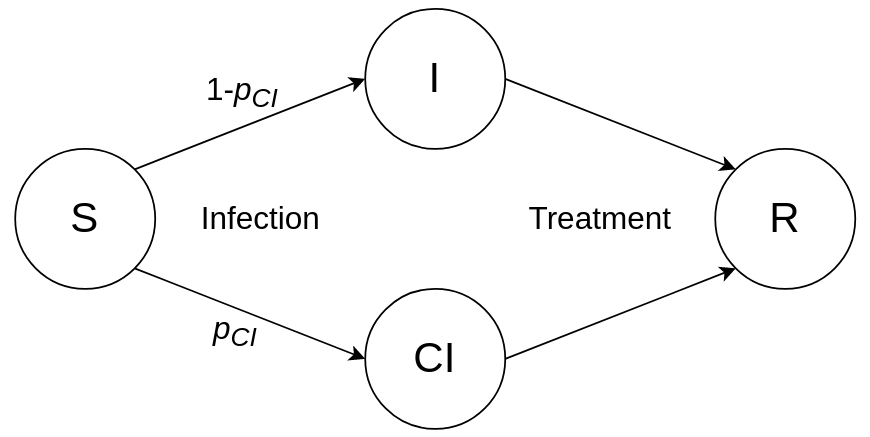}}
\caption{Adapted epidemiological model}
\label{fig}
\end{figure}

The main parameter is the infection rate $\lambda$ for a contact link in the network. The infection rate depends on the intensity of the contact, its type (e.g., with or without mask), and the contagiousness of the virus itself. Since the virus can affect people in different ways depending on their gender and age, as well as depending on comorbidities and other factors, in order to take into account these features in the model of disease spread in Ukraine, it was decided to use the demographic distribution as of 2020. (Age structure of the population of Ukraine \url{https://www.lv.ukrstat.gov.ua/dem/piramid/all.php}). Data on the number of infected individuals and mortality as of December 16, 2020, were also used, see operational monitoring of the situation around COVID-19: \url{https://nszu.gov.ua/e-data/dashboard/covid19, retrieved Dec. 2020}.

\subsection{Models of the topology of contacts in a social network}
The dynamics of viral spread depends on the network topology, including local characteristics (e.g., local clustering and degrees of nodes) and global characteristics (e.g., eigenvalue spectrum, shortest path characteristics). 
%
Real social networks do not have a clear structure but may have certain patterns. Therefore, to describe them approximately it is common practice to use random graph models of complex networks, where networks are generated according to certain rules and probability distributions. Within a random graph model, certain properties of networks are typically shared, such as small world or clustering characteristics, and so on.
Networks of human contacts are displayed in the form of graphs where an edge connects individuals (nodes) that are in contact with each other. Because human interaction is often bidirectional, we consider here undirected graphs, noting that the simulation methods in this paper can be easily adapted to directed graphs. We investigate the dynamics on a set of typical network topologies represented by random graph models. Each of the topologies discussed here is described in detail in~\cite{barabasi2013network}.

The most common network topology discussed in classical epidemiology is that of a \emph{Complete Graph}(CG), where each node is connected to every other node. However, in real networks, other topologies are more common, such as small-world networks, and scale-free networks. Small-world models assume a small graph distance between people in a social network (an example is a rule of "6 handshakes").
Scale-free networks are closest to real networks, including social contact networks. Scale-free networks are subject to the power law, where the probability  $P(k)$ of the degree $k$ of the nodes follows the law
$P (k) {\displaystyle \sim } k^{- \gamma}$. 

\subsubsection{Erdős–Rényi model}
The Erdős–Rényi graph model(ER) is a random graph where $L$ links are randomly distributed across a set of $N$ nodes\footnote{There is an alternative formulation of ER graph models in which links are retained with a probability $p$. We do not use this formulation in order to keep the number of edges constant.}. The degrees of nodes in an ER graph follows a binomial distribution, and not a power law. It is also known, that for $L > N \log N$ the network tends to be fully connected, whereas if $L < N$ the network tends to be fragmented into many small isolated components.

\subsubsection{Watts-Strogatz model}
The Watts-Strogatz model(WS) is a small-world model. The construction of the Watts-Strogatz model begins with a grid in which each node is connected strictly to $m$ neighbors, after which each of the edges can reconnect to a randomly selected node with a probability $p$, this process is called \emph{reconnection}. As a result of reconnection, the average distance between nodes decreases.
The Watts-Strogatz model is characterized by high clustering, and also by a small average shortest path $l$ which decreases with increasing probability of reconnection of node $p$ .

\subsubsection{Barabási–Albert model}
The Barabási–Albert model(BA) is a scale-free network with exponent $\gamma = 3$, and it is built on the principle of growth and preferential attachment, that is, at each step a node with $m$ edges is added. New nodes are linked to others by \emph{preferential attachment} to nodes with a probability that is proportional to the current degree of the other node.

%
The preferential attachment process leads to the creation of \emph{hubs}, i.\,e. a few nodes to which are connected many other nodes with, on average, smaller degrees. These hubs serve as shortcuts for information or diseases spreading through the networks. Therefore the average shortest distance in such networks tends to be small.

\begin{@twocolumnfalse}
\begin{figure}
\centering
    \begin{subfigure}{.24\textwidth}
      \includegraphics[width=\linewidth]{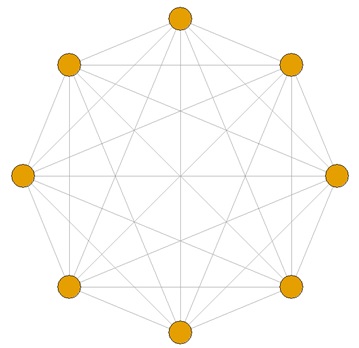}
      \caption{Complete graph for n=8}
      \label{fig:sub_first1}
    \end{subfigure}
    \begin{subfigure}{.24\textwidth}
      \includegraphics[width=\linewidth]{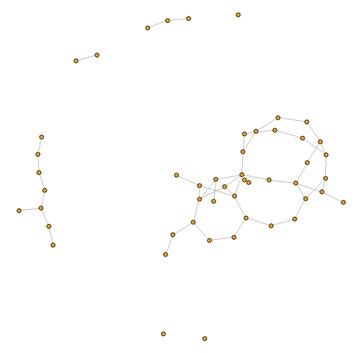}
      \caption{Erdős–Rényi model for $N = 50$, $L = 50$}
      \label{fig:sub_second1}
    \end{subfigure}
    \\
    \begin{subfigure}{.24\textwidth}
      \includegraphics[width=\linewidth]{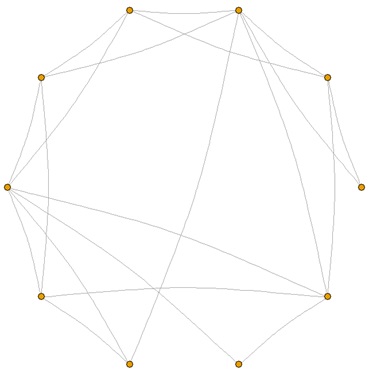}
      \caption{Watts-Strogatz model for N=10, m=4, p=0.1}
      \label{fig:sub-third1}
    \end{subfigure}
    \begin{subfigure}{.24\textwidth}
      \includegraphics[width=\linewidth]{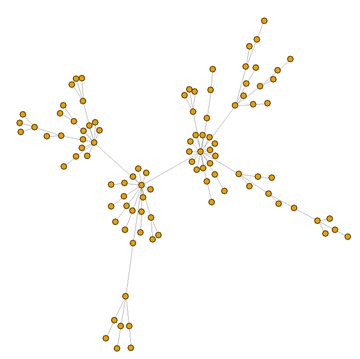}
      \caption{Barabási–Albert model for N=50, m=1}
      \label{fig:sub-fourth1}
    \end{subfigure}
\caption{Network models}
\label{fig:fig1}
\end{figure}

\end{@twocolumnfalse}

\subsection{Infection Model and Stochastic Simulation Algorithm}
For modeling the infection process in this work we use Continuous Time Markov Chains (CTMCs)\cite{norris_1997CTMC}. In contrast to other models such as discrete Markov chains and cellular automata, CTMCs feature realistic modeling of time. Moreover, they are not based on asymptotical simplifications and stability of mean values as do the classical epidemiological models based on differential equations and, respectively, mean field models. To tame the state-space explosion we make use of the underlying principles of Gillespie's stochastic simulation algorithm: (1) simulation of time between two state transitions and the simulation of the next state can be separated, and (2) only a small number - linear in the number of nodes - of state transitions can occur in a single step of the simulation~\cite{gillespie2007stochastic}.

In the CTMC model of the contact process, all rates at which transitions occur between network states, is described in a generator matrix $Q$.
This matrix is a $2^N \times 2^N$ matrix, where N denotes the number of nodes in the contact network.  In the matrix a component  $q_{ij}$ describes  the rate at which the system changes from network state $j$ given it is currently in state $i$, and $q_{ii} = -\sum_{i\neq j} q_{ij}$ is the rate of leaving state $i$ (on the diagonal). Using a state space of size $2^N$ becomes however computationally prohibitive for larger $N$, and we next propose a way of avoiding the representation of the full generator matrix.

In the specific case of simulating the SIR epidemic process, the only non-zero transition rates are those where a single additional node gets infected or where a node that is infected is removed from the network. Let $i$ denote the current state and $i_{+j}$  denote a network state where node $j\in[1,N]$ is the index of a susceptible node that potentially gets infected in addition to the previously infected nodes in state $i$. Then the non diagonal state transitions $q_{ij}$, $i\neq j$ are computed as follows:

\begin{equation}
q_{i,i_{+j}} =
\begin{cases}
  \lambda v(j), \text{ if node $j$ is susceptible;}\\
  0, \text{if node $j$ is not susceptible.}
\end{cases}
\end{equation}
Here $\lambda$  is the infection rate of the virus, $v(j)$ is the number of infected neighbors of the vertex $j$ in state $i$.
The transition time from the state is exponentially distributed, with probability density function $F$ defined as:
\begin{equation}
 F(x) =
\begin{cases}
  \theta e^{-\theta x}  \mbox{ if } x \geq 0;\\
  0, \mbox{ otherwise}\\
\end{cases}
\end{equation}
where $\theta = q_{ii}$. The expected value of the time until the next infection event is given by $1/\theta$ for a single link, and in a given state of the entire network in the CTMC by $\frac{1}{\lambda q_{ii}}$. 
%
%
\begin{figure}[h]
\centering
\includegraphics[width=0.5\textwidth]{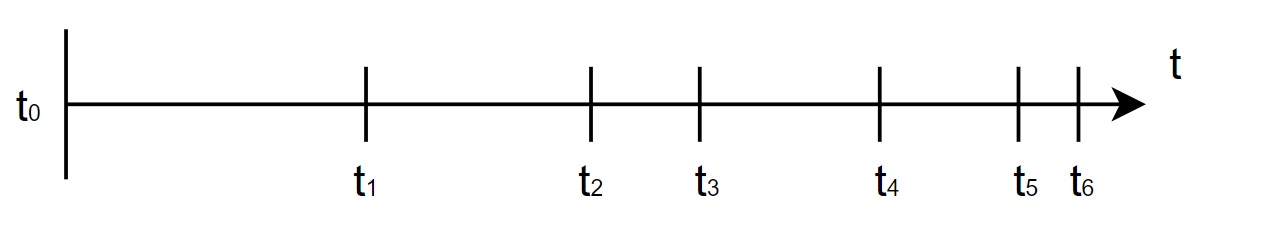}
\caption{Example of expected time}
\label{fig:time1}
\end{figure}
The probability of transition from state $i$ to state $j$, that is $p_{ij}$ is determined by the formula:
\begin{equation}
p_{ij} =  \frac{q_{ij}}{\sum_{l, l \neq i} q_{il}}
\end{equation}
and it can be simulated by “roulette wheel” simulation~\cite{roulette_wheel}. 


\subsubsection{Our implementation}
For the implementation of Gillespie's algorithm, we use matrix multiplication using the \texttt{Numpy} library for \texttt{Python} to make the calculation efficient.
We represent the network as an $N\times N$ adjacency matrix $A$, with $a_{ij} = 1$ if there is an edge between the nodes, and $a_{ij} = 0$ otherwise. We preserve the state of the node using two vectors: the susceptibility state vector $\mathbf{s}$, in which $s_i = 1$ if the node $i$ is susceptible and 0 if not susceptible, and also, the contagious state vector $\mathbf{c}$, in which $c_i = 1$ if the node $i$ is contagious and 0 if not contagious. In each iteration (single event simulation) the algorithm determines the single node which will infect at the time of the next event. The purpose of this strategy is to also compute on-the-fly the  number of reproductions of the virus $R$ per node or individual. An alternative would be to determine first which single additional node gets infected next, which was implemented in a previous work by the authors \cite{kuryliak2021effect}.

\textbf{Iteration 0}

Step 0: Infecting a randomly chosen node (virus enters the network).
Setting $c_i = 1$, $s_i = 0$, where $i$ - index of chosen node.

Step 1: Calculation of number of contacts with susceptible nodes for all nodes by multiplication of contact network's adjacency matrix $A$ by a transposed susceptibility state vector $\mathbf{s}^T$.
\begin{equation}
    \label{eq:contacts_S_ALL}
    A \cdot \mathbf{s}^T
\end{equation}

Step 2: Selection of number of contacts with susceptible nodes only for contagious nodes by multiplication the transposed result of the Step 1 step by vector of contagious nodes $\mathbf{c}$. 
\begin{equation}
    \label{eq:contacts_S_C}
    (A \cdot \mathbf{s}^T)^T \circ \mathbf{c}
\end{equation}

Step 3: Computation of total number of contacts of contagious nodes $T$ with nodes that are able to infect  at next time step(contagious) as the sum of results of Step 2.
\begin{equation}
    \label{eq:total_contacts_S_С}
    T = \sum n, \forall n \in (A \cdot \mathbf{s}^T)^T \circ \mathbf{c}
\end{equation}

Step 4: Calculation of time to the next infection event by sampling from an exponential distribution. \begin{equation}
    \label{eq:time_to_event}
    \Delta t = \theta e^{-\theta x}\text{,}
\end{equation}
where $\theta = \lambda T$.

\textbf{Further iterations $(1, 2, ...)$:}

Step 0: Recovering nodes. Setting $s_i = 0$ and $c_i = 0$, for nodes which were previously infected but due to the time passed are no longer infectious (recovered (immune), quarantined, or dead).

Step 1: Calculation of number of contacts with susceptible nodes for all nodes by formula \ref{eq:contacts_S_ALL}

Step 2: Selection of number of contacts with susceptible nodes only for contagious nodes by formula \ref{eq:contacts_S_C}

Step 3: Compute total number $T$ of contacts of contagious nodes with susceptible by formula \ref{eq:total_contacts_S_С}

Step 4: Dividing of the vector found on Step 2 by $T$ to find probabilities for all nodes to infect (do infection) at the next event.
\begin{equation}
    \mathbf{p} = \dfrac{(A \cdot \mathbf{s}^T)^T \circ \mathbf{c}}{T}
\end{equation}

Step 5: Choosing a node which will infect, proportionally to probabilities of the vector $\mathbf{p}$ (by "roulette wheel").

Step 6: Choosing a neighbor of the node which will be infected using uniformly distributed probabilities.

Step 7: Recalculation of total number of contact of contagious nodes with susceptible nodes (repeating of steps 1-3)

Step 8: Calculation of time to next infection by formyla \ref{eq:time_to_event} and adding it to the total time that has passed.

Steps 1-8 are repetting until the there are no nodes that can be infected or simulation time is over.

Data on number nodes in each state is collected at equidistant points in time (end of a day), which we denote as "day".

\subsubsection{List of input parameters}
\begin{enumerate}[wide, labelwidth=!, labelindent=0pt]
    \item [$\bullet$] Network size and link density;
    \item [$\bullet$] Network model (Erdős–Rényi, Barabási–Albert, Watts-Strogatz, complete graph);
    \item [$\bullet$] Infection rate (contagiousness);
    \item [$\bullet$] Intensity of contacts (link weight, the same for all contacts was used for the paper);
    \item [$\bullet$] Demographic distribution (age, gender), in proportion to which individuals are generated in the network;
    \item [$\bullet$] A distribution of the probability of critical infections;
    \item [$\bullet$] Term of being in infected  and critically infected states; (How long the individual will be in this state).
\end{enumerate}

\subsection{Software specification}
The largest unit of data over which the action is performed is the adjacency matrix of dimension $N\times N$, where $N$ is the number of nodes in the network, therefore, memory usage is proportional to $N^{2}$. The running time of the program to infect all nodes of the network is proportional to $N^{3}$, because the most difficult part is finding the number of infected neighbors, which is implemented by multiplying the adjacency matrix by the vector column of contagious nodes, and it is done $N$ times to infect all nodes.

\begin{table}[ht]
\caption{
\label{table:exp21} Resources} 
\centering 
\begin{tabular}{c c c}
    \toprule 
        {Number nodes} &
        {RAM (MiB)} &
        {Time (second)} \\ [0.5ex] 
    \midrule 
    200 & 86 & 0.085 \\
    500 & 92 & 0.72 \\
    1000 & 129 & 4.27 \\
    2000 & 200 & 29.5 \\
    5000 & 280 & 360 \\
    10000 & 850 & 2100 \\
    \bottomrule 
\end{tabular} 
\end{table}

Table 1 shows time and memory usage for different number nodes in the network. Data on the operation of the program was obtained using an AMD Ryzen 5 3500U processor, Python version 3.9.7, OS Fedora 34 kernel-5.14.18-200. The arithmetic mean of time was taken from 10 simulations for the complete graph in which all nodes were infected. All simulations were performed in single-threaded mode.

The link to the backend part (algorithm without visualization) is
\url{https://github.com/YulianKuryliak/VirusSpreadingSimulation}.\\
The link to the dashboard is
\url{https://github.com/YulianKuryliak/EpidemicOutbreakPredictor}.

\section{Results}

\subsection{Study of the influence of individual and weighted probability of critical infection}

To understand the difference of using individually different vs. a constant probability of getting critically infected we made experiments on complete graphs to eliminate the influence of network topology. A network with 10000 nodes was used for this first study. We computed median values for each point of time from 15 simulations with individual and weighted infection. The simulations were completed on networks with 10000 nodes and an infection rate of 0.00005. In the experiments, 50 days were recorded, starting from the first occurrence of the virus in the contact network.

In this and further studies, the demographic distribution of nodes is proportional to the demographic distribution in 2020 of Ukraine. 
As the value of the weighted probability of getting critically infected we used the weighted probability of death $p_{d}$ increased by 0.2. Here $p_{d}$  is  is calculated by the weighted average formula
\begin{equation}
    p_{d} = \frac{\sum_{i}n_{m i} * p_{m i} + n_{w i} * p_{w i}}{n},
\end{equation}
where
$n_{m i}$ - number of men in age range,
$n_{w i}$ - number of women in age range,
$p_{m i}$ - probability of death for men of age range,
$p_{w i}$ - probability of death for women of age range,
$n$ -  total number of people,
$i$ - age range.

\begin{figure}
\centering
\includegraphics[width=\columnwidth]{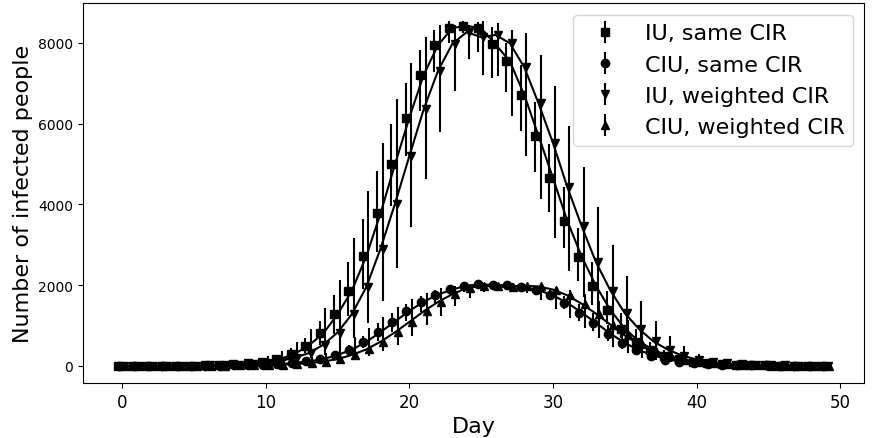}
\caption{Comparison of simulations with the weighted and individual probabilities of critical infection (CI), with an error deviation of ± 25\% from the median values.}
\label{fig:exp1}
\end{figure}

The correlation between ICUs ('intensive care units') and total number of infected nodes in each point of time from the simulations are shown in Tables 2 and 3, respectively. According to the results, slightly higher values of the dependency is observed in simulations with different infection rate on the intervals in range 10-40 days.

For the simulations with the weighted infection rate the PIN('peak number of infected') is observing on 24 day (8266 nodes) and PCIN('peak number of critically infected nodes') is observing on 28 day(1985 nodes).
For the simulations with individual infection rate the PIN is observing on 24 day (8416 nodes) and PCIN is observing on 25 day(2017 nodes).

As is shown in Fig.\ref{fig:exp1} and tables \ref{table:ICU_dep_individual} and \ref{table:ICU_dep_weighted}, the number of ICUs has a strong correlation with a number of infected nodes, but not proportional, because of different term of being in states. The PCIN is observing after the PIN and correlation between number of ICUs and number of infected nodes is lower after the PIN of the outbreak is observing and fraction of ICUs increases in time. Therefore, hospitals have to be ready for the peak load after the PIN is observed and the high load in the end of an outbreak.

\begin{table}[htb]
    \caption{
    \label{table:ICU_dep_individual} The dependency values of ICU from total number of infected for individual infection rate. Fraction means fraction of ICU divided by total number of infected nodes.} 
    \centering 
    \begin{tabularx}{0.49\textwidth} { 
      >{\centering\arraybackslash}X 
      >{\centering\arraybackslash}X 
      >{\centering\arraybackslash}X 
      >{\centering\arraybackslash}X 
      }
     \toprule
     Interval of days & Median fraction & Average fraction & Correlation coefficient \\
     \midrule
        0-50 & 0.257 & 0.342 & 0.96 \\
        0-24 & 0.221 & 0.192 & 1.0 \\
        10-24 & 0.221 & 0.226 & 1.0 \\
        10-40 & 0.252 & 0.347 & 0.938 \\
        15-35 & 0.239 & 0.3 & 0.867 \\
        24-40 & 0.461 & 0.452 & 0.946 \\
        24-50 & 0.546 & 0.481 & 0.959 \\
    \bottomrule
    \end{tabularx}
    \end{table}

\begin{table}[htb]
    \caption{
    \label{table:ICU_dep_weighted} Fraction of individuals in ICU (critically infected nodes/ total number of infected nodes) vs. infection rate} 
    \centering 
    \begin{tabularx}{0.49\textwidth} { 
      >{\centering\arraybackslash}X 
      >{\centering\arraybackslash}X 
      >{\centering\arraybackslash}X 
      >{\centering\arraybackslash}X 
      }
    \toprule
     Interval of days & Median fraction & Average fraction & Correlation coefficient \\
    \midrule
    0-50 & 0.297 & 0.358 & 0.955 \\
    0-24 & 0.215 & 0.216 & 1.0 \\
    10-24 & 0.21 & 0.217 & 0.999 \\
    10-40 & 0.243 & 0.328 & 0.93 \\
    15-35 & 0.237 & 0.283 & 0.863 \\
    24-40 & 0.42 & 0.424 & 0.937 \\
    24-50 & 0.554 & 0.489 & 0.955 \\ 
    \bottomrule 
    \end{tabularx} 
\end{table}



\subsection{Scaling of the number of infected nodes}
As usual, epidemic outbreaks occur in large networks and there are not always methods and computational resources to predict consequences in the whole network. Therefore we do the study of the possibility of the scalling of the fraction of infectied nodes from all nodes.
For the study of the effect of scaling where chosen Erdős–Rényi of the size of $200$, $500$, $1000$, $2000$ and $10000$ of nodes and average degree of a node $4$ for all cases.

\begin{figure}[h]
\centering
\includegraphics[width=\columnwidth]{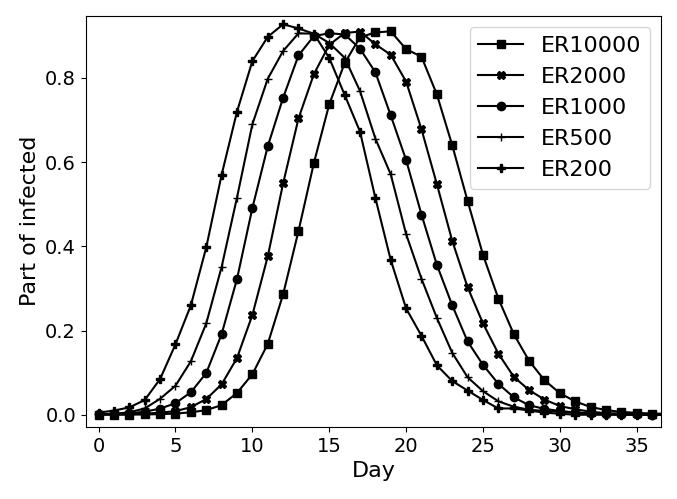}
\caption{Scaling of number of infected in networks of different sizes}
\label{fig:scalling}
\end{figure}

According to Fig.\ref{fig:scalling}, the results of PIN/PCIN can be scalable, and the fraction of PIN is about the same for networks of all sizes, but the time of the PIN in bigger networks is farther than in smaller ones. Thus, prediction of PIN using scaling is possible, but it does not give accurate information about the time of the peak and how many nodes are infected at each point of time. 


\subsection{Study of the influence of network topology on the number of simultaneously infected nodes}
For the study of the influence of network topology on the peak number of infected nodes, the most common models of complex networks were chosen, such as Erdős–Rényi random graph, a small-world network model Watts–Strogatz, and a scale-free network model Barabási–Albert were used. Were used three different reconnection rates(0.1, 0.2, and 0.5) for the Wats-Strogars model with the purpose to have networks with various properties for the same number of nodes and average degree of a node. All networks consist of 10000 nodes and 4 initial edges for a node. Properties of the networks are shown in table \ref{table:exp21}.

\begin{table}
    \caption{
    \label{table:exp21} Parameters of the network} 
    \centering 
    \begin{tabularx}{0.49\textwidth} { 
      c
      >{\centering\arraybackslash}X 
      >{\centering\arraybackslash}X 
      >{\centering\arraybackslash}X 
      >{\centering\arraybackslash}X 
      >{\centering\arraybackslash}X 
      >{\centering\arraybackslash}X 
      }
     \toprule
        \thead{Network \\ model} &
        {CCG} & 
        {NACC} & 
        {ADN} & 
        {MDN} & 
        {D} & 
        {ASPL} 
        \\
     \midrule
        ER & 0.001 & 0.001 & 8.003 & 8.0 & 8.2 & 4.663 \\
        BA & 0.004 & 0.005 & 7.998 & 5.7 & 6.0 & 4.004 \\
        WS 0.1 & 0.333 & 0.349 & 8.0 & 8.0 & 9.0 & 5.929 \\
        WS 0.2 & 0.159 & 0.173 & 8.0 & 8.0 & 8.1 & 5.177 \\
        WS 0.5 & 0.01 & 0.011 & 8.001 & 8.0 & 8.1 & 4.708 \\ [0.1ex]
    \bottomrule
    \end{tabularx}
      CCG - Clustering coefficient (global), NACC - Network average clustering coefficient, ADN - Average degree of node, MDN - Median degree of node, D - Diameter of network, ASPL - Average shortest path length.\\
      *The values in this and the following tables are averaged for a sample of 100 networks.
\end{table}

\begin{figure}
\centering
\begin{subfigure}{0.5\textwidth}
      \includegraphics[width=\columnwidth]{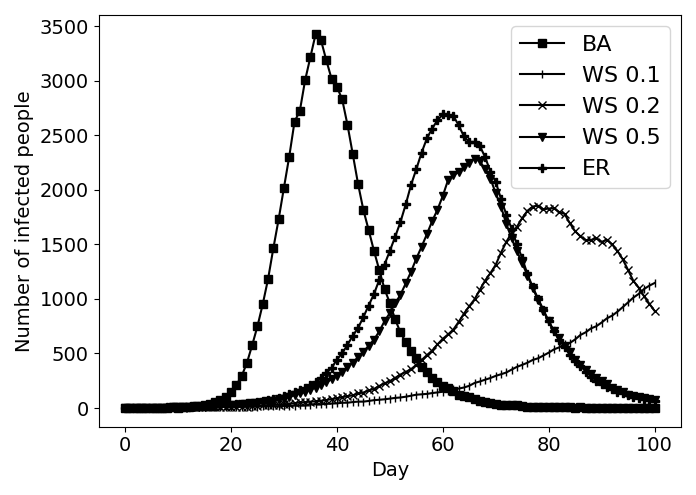}
\caption{Plots for infection rate $\lambda = 0.025$}
\label{fig:network_comparition01}
\end{subfigure}

\begin{subfigure}{0.5\textwidth}
\includegraphics[width=\columnwidth]{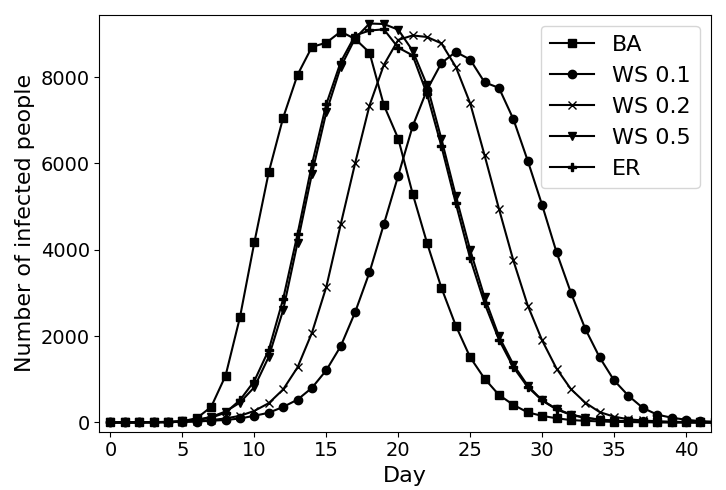}
\caption{Plots for infection rate $\lambda = 0.1$}
\label{fig:network_comparition0025}
\end{subfigure}
\caption{Plots of number of infected nodes at each point of time}
\label{fig:network_comparition}
\end{figure}

The median values of infected nodes at each point of time in all networks are shown in Fig. \ref{fig:network_comparition}.
According to the results of simulations for infection rate $\lambda = 0.025$, there is a difference of PINs: the longer is an average shortest path the smaller is a PIN and farther is the time of the peak.
According to the results of simulations for infection rate $\lambda = 0.1$, the peak number of infected is about the same, but the time of the peak is still farther for networks with a longer average shortest path. PIN is lower in BA network than in ER and WS with the reconnection rate of 0.5 due to very short average shortest path length, because of a small number of concentrators with very high degree. Thus, concentrators infect quickly after an outbreak is started, and recover faster than all their neighbors are infected. The median degree of nodes of BA network is lower, therefore, after removing a concentrator the average shortest path length significantly increases. In spite of that, random networks contain more concentrators with lower degrees.
So, for the middle infection rate in networks with high concentrators degree, the collapsing is observed, but it does not qualitatively change the peak number of infected. Correlation values are given in tables \ref{table:CorrPIN} and \ref{table:CorrTime}

\begin{table}[htb]
    \caption{
    \label{table:CorrPIN} Correlation coefficients of the number peak infected with network properties} 
    \centering 
    \begin{tabular}{cccccc}
     \toprule
        {IR} &
        {CCG} & 
        {NACC} & 
        {MDN} & 
        {ASPL} & 
        {Time}
        \\ 
    \midrule
    0.025 & -0.88 & -0.881 & -0.747 & -0.985 & -0.996\\
    0.1 & -0.942 & -0.939 & -0.142 & -0.787 & -0.833\\
    0.2 & -0.805 & -0.795 & -0.459 & -0.838 & -0.809\\
    \bottomrule
    \end{tabular}
    Time - a day when PIN is observed, IR - infection rate.
\end{table}

\begin{table}[htb]
    \caption{
    \label{table:CorrTime} Correlation coefficients of time of the peak with network properties} 
    \centering 
    \begin{tabularx}{0.49\textwidth}{c c c c c}
     \toprule
        {IR} &
        {CCG} & 
        {NACC} & 
        {MDN} & 
        {ASPL}  
        \\ 
    \midrule 
    0.025 & 0.881 & 0.882& 0.76 & 0.993\\
    0.1 & 0.932 & 0.933 & 0.66 & 0.988\\
    0.2 & 0.655 & 0.651 & 0.829 & 0.866\\
    \bottomrule
    \end{tabularx}
\end{table}

\begin{figure}[h]
\centering
    \begin{subfigure}{.5\textwidth}
    \includegraphics[width=\columnwidth]{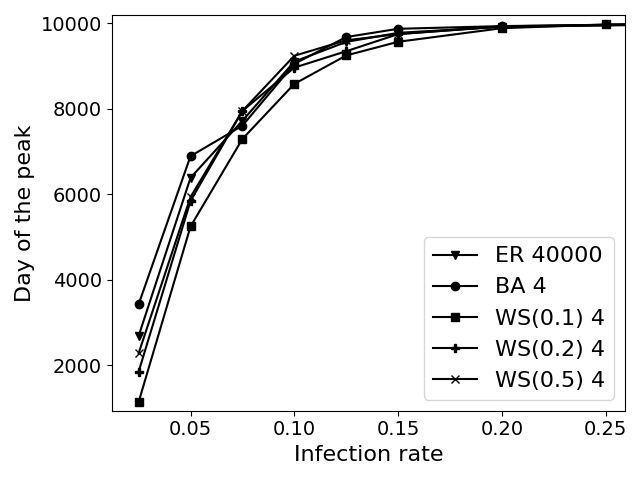}
    \caption{Plot of peak numbers of infected}
    \label{fig:exp2_PIN}
    \end{subfigure}
    
    \begin{subfigure}{.5\textwidth}
    \includegraphics[width=\columnwidth]{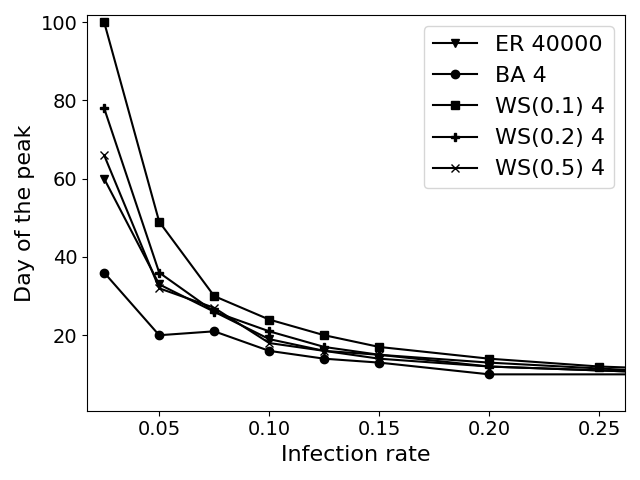}
    \caption{Plot of the time of peak}
    \label{fig:exp2_PIN_time}
    \end{subfigure}
\caption{Plots of peak values for different network topologies depending on the infection rate}
\label{fig:exp2_peak}
\end{figure}

As is shown in Fig. \ref{fig:exp2_peak}, network topology has a significant effect on the peak values and time of the peak for small infection rate, but the significance of a topology decreases with increasing the infection rate of a virus. Firstly, the significance of the topology on the number of infected is lost, and then at the time of this peak. Anyway, it is possible to postpone the peak number of infected increases average shortest path length even for high infection rate.


\section{Conclusion}
An epidemic outbreak has been simulated in complex networks generated according to models such as the Erdős–Rényi random model, the Watts-Strogatz small-world model, the Barabási–Albert scale-free model, and the complete graph using continuous time Markov chains for which we propose an efficient stochastic simulation technique (based on Gillespie's algorithm) for computing their stochastic trajectories within the compartemental mod.
 of type SIR. The simulation time scales proportionally to the number of nodes in the network, and thus avoids the state space explosion observed for general CTMC on the given network state space, without compromising for accuracy of the simulation. The simulation algorithm can be used to simulate efficiently the trajectory of networks with a number of up to 10000 individuals on conventional hardware.
 
In the second part of the paper we use the simulation to study the influence of network topology on networks with the same number of average contacts per node. Special attention is on peak number of individuals requiring intensive care, and when this peak occurs (early, delayed). This has been motivated by the practical importance of these indicators, when minimizing the risk of running out of capacity in terms of available ICUs and, respectively, of having sufficient time to prepare for the critical situation. 

It has been found that the number of infected at the same time(ie, peak values) and, consequently, the number of critically infected people, depends on the rate of virus transmission on the network and has a strong correlation with \emph{average shortest path length}, but not to a number of other graph features - clustering coefficient, average degree - for which it was computed. The rate of virus spread changes at the same infection rate for different network models with the same average node degree. The virus was found to spread fastest in the Barabási–Albert scale-free network model, slower in the Erdős–Rényi random network model, and slowest in the Watts-Strogatz small-world network model. 

As the ICUs seems to critically determine the workload of hospitals, it is important to take targeted testing and quarantine measures to increase the average path length and thereby to reduce the height of this peak (PCIN). If contagiousness of a virus is high, decreasing of the \emph{average shortest path} length has a relative small effect, but such measure delays the time of the peak.

Since the chosen term of being in critically infected state is longer than in the simply infected state, PCIN is observed after PIN, and the fraction of ICUs grows until the end of the pandemic.

Also, was shown that results of peak number of infected nodes can be scalable for the same degree of a node (about the same part of nodes is infected for networks with different amount of nodes), but time of the peak is the shorter the smaller is a network.

In addition to these findings, the importance of taking into account individual node characteristics has been analysed, and our empirical evidence shows that for the scenarios we analyzed (Ukraine data set, networks of size 10000) they seem to eb not very important for large networks and weighted values can be used, but in small networks, it is difficult to maintain demographic distribution, so it is better to apply individual node characteristics when they are important. Also, is shown that peak load on hospitals is after peak number of infected nodes due to longer term of being critically infected.



\section{Outlook}
This study makes some interesting contributions to the modelling of an epidemic, and in particular the COVID-19 pandemic, focusing on the aspects of network topology and demographics. However there are still many possible enhancements that can be implemented for this simulation model in future work:
\begin{enumerate}
\item We use the SIR model, but it does not pay attention for undetected infected people and those which do not develop symptoms. There is however to our knowledge little reliable data for modeling these effects.
\item We do not yet model the influence of testing, that is increasingly used to monitor people, and allows also to detect and isolate infectious people who do not have symptoms.
\item We do not yet model the effect of different mutants of viruses on the dynamics. It is known, that the contagiousness and the resistance to certain vaccines is different for different mutations of the virus. The simulation of different virus strands can be elegantly incorporated in a CTMC model by extending the state space from the space of binary to the space of $k$-ary vectors (each positive integer representing a different virus type), but would require an analysis on its own.
\item We do not yet use any immunisation strategy and non-pharmaceutical interventions (like in papers \cite{ferguson2006strategies, halloran2008modeling}). The results of this study might however hint at a reconsideration of the average shortest path length as an important indicator of network vulnerability - to be increased by means of contact restrictions.
\item The models are for small to moderate size scale contact networks and they model the outbreak on the individual level. To scale up the modelling to the level of countries will be challenging due to the large number of individuals and networks and a promising route would be to use hierarchical network models that model the interaction between regions~\cite{ACHTERBERG2020}, but using simulations of contact networks instead of empirical approximation for each region.
\end{enumerate}

\section{Acknowledgements}
Michael Emmerich acknowledges funding from the European Union’s Horizon 2020 research and innovation programme under the Marie Skłodowska-Curie grant agreement Nr. 823866.

\bibliographystyle{unsrtnat}
\bibliography{references}  






\begin{appendices}
\section{An example for SI model on complete graph}

Adjacency matrix
\[
A = 
\begin{pmatrix}
1 & 1 & 1 & 1\\
1 & 1 & 1 & 1\\
1 & 1 & 1 & 1\\
1 & 1 & 1 & 1\\
\end{pmatrix}
\]

Vector of susceptible nodes $\mathbf{s}$
\[
\mathbf{s} = 
\begin{pmatrix}
0 & 1 & 1 & 1\\
\end{pmatrix}
\]

Vector of contagious nodes $\mathbf{c}$
\[
\mathbf{c} =
\begin{pmatrix}
1 & 0 & 0 & 0\\
\end{pmatrix}
\]

Iteration 0

Step 1: Multiplication of adjacency matrix $A$ by the vector of susceptible nodes $\mathbf{s}$ with the purpose to calculate the number of contacts with susceptible nodes for all nodes.

\[
A \cdot \mathbf{s}^T = 
\begin{pmatrix}
1 & 1 & 1 & 1\\
1 & 1 & 1 & 1\\
1 & 1 & 1 & 1\\
1 & 1 & 1 & 1\\
\end{pmatrix}
\cdot
\begin{pmatrix}
    0\\
    1\\
    1\\
    1\\
\end{pmatrix}
=
\begin{pmatrix}
    3\\
    3\\
    3\\
    3\\
\end{pmatrix}
\]

Step 2: Multiplication the result of the previous step by vector of contagious nodes $\mathbf{c}$. We do it with the purpose to concentrate only on number of contacts with susceptible nodes for nodes that are able to infect at next time step.


$$
(A \cdot \mathbf{s}^T)^T \circ \mathbf{c} = 
\begin{pmatrix}
3 & 3 & 3 & 3\\
\end{pmatrix}
\circ
\begin{pmatrix}
1 & 0 & 0 & 0\\
\end{pmatrix}=
\centering
\begin{pmatrix}
3 & 0 & 0 & 0\\
\end{pmatrix}
$$

Step 3: Compute total number $T$ of contacts of contagious nodes with susceptible as the sum of results of Step 2.
\[
T = \sum ((A \cdot \mathbf{s}^T)^T \circ \mathbf{c}) = 3
\]

Step 8: Calculation of time to next infection using exponential distribution $\Delta t = \theta e^{-\theta x}$ ,where $\theta = T$

\[
\Delta t = 1/3
\]

Iteration 1

Step 1: Multiplication of adjacency matrix $A$ by the vector of susceptible nodes $\mathbf{s}$ with the purpose to calculate the number of contacts with susceptible nodes for all nodes.

\[
A \cdot \mathbf{s}^T = 
\begin{pmatrix}
1 & 1 & 1 & 1\\
1 & 1 & 1 & 1\\
1 & 1 & 1 & 1\\
1 & 1 & 1 & 1\\
\end{pmatrix}
\cdot
\begin{pmatrix}
0\\
1\\
1\\
1\\
\end{pmatrix}
=
\begin{pmatrix}
3\\
3\\
3\\
3\\
\end{pmatrix}
\]

Step 2: Multiplication the result of the previous step by vector of contagious nodes $\mathbf{c}$. We do it with the purpose to concentrate only on number of contacts with susceptible nodes for nodes that are able to infect at next time step.


$$
(A \cdot \mathbf{s}^T)^T \circ \mathbf{c} = 
\begin{pmatrix}
3 & 3 & 3 & 3\\
\end{pmatrix}
\circ
\begin{pmatrix}
1 & 0 & 0 & 0\\
\end{pmatrix}
=
\centering
\begin{pmatrix}
3 & 0 & 0 & 0\\
\end{pmatrix}
$$
Step 3: Compute total number $T$ of contacts of contagious nodes with susceptible as the sum of results of Step 2.
$$
T = \sum ((A \cdot \mathbf{s}^T)^T \circ \mathbf{c}) = 3
$$

Step 4: Dividing of the vector found on step 2 by $T$ to find probabilities for all nodes to infect at the next event.
$$
\begin{pmatrix}
3 & 0 & 0 & 0\\
\end{pmatrix}
/ 3 = 
\begin{pmatrix}
1 & 0 & 0 & 0\\
\end{pmatrix}
$$

Step 5: Choosing a node which will infect.

node 1 will infect

Step 6: Choosing a neighbor of the node which will be infected.

p = (0, 1/3, 1/3, 1/3)

lets choose node 2

Step 7: Recalculation of total number of contact of contagious nodes with susceptible (repeating of steps 1-3)

$$
A \cdot \mathbf{s}^T = 
\begin{pmatrix}
1 & 1 & 1 & 1\\
1 & 1 & 1 & 1\\
1 & 1 & 1 & 1\\
1 & 1 & 1 & 1\\
\end{pmatrix}
\cdot
\begin{pmatrix}
0\\
0\\
1\\
1\\
\end{pmatrix}
=
\begin{pmatrix}
2\\ 
2\\ 
2\\ 
2\\
\end{pmatrix}
$$

$$
(A \cdot \mathbf{s}^T)^T \circ \mathbf{c} = 
\begin{pmatrix}
2 & 2 & 2 & 2\\
\end{pmatrix}
\circ
\begin{pmatrix}
1 & 1 & 0 & 0\\
\end{pmatrix}
=
\centering
\begin{pmatrix}
2 & 2 & 0 & 0\\
\end{pmatrix}
$$
$$
T = \sum ((A \cdot \mathbf{s}^T)^T \circ \mathbf{c}) = 4
$$

Step 8: Calculation of time to next infection using exponential distribution $\Delta t = \theta e^{-\theta x}$ ,where $\theta = T$

$$
\Delta t = 1/4
$$

Iteration 2

Step 1: 
$$
A \cdot \mathbf{s}^T = 
\begin{pmatrix}
1 & 1 & 1 & 1\\
1 & 1 & 1 & 1\\
1 & 1 & 1 & 1\\
1 & 1 & 1 & 1\\
\end{pmatrix}
\cdot
\begin{pmatrix}
0\\
0\\
1\\
1\\
\end{pmatrix}
=
\begin{pmatrix}
2\\ 
2\\ 
2\\ 
2\\
\end{pmatrix}
$$
Step 2:
$$
(A \cdot \mathbf{s}^T)^T \circ \mathbf{c} = 
\begin{pmatrix}
2 & 2 & 2 & 2\\
\end{pmatrix}
\circ
\begin{pmatrix}
1 & 1 & 0 & 0\\
\end{pmatrix}
=
\centering
\begin{pmatrix}
2 & 2 & 0 & 0
\end{pmatrix}
$$
Step 3:
$$
T = \sum ((A \cdot \mathbf{s}^T)^T \circ \mathbf{c}) = 4
$$
Step 4: 
$$
\begin{pmatrix}
2 & 2 & 0 & 0\\
\end{pmatrix}
/ 4 = 
\begin{pmatrix}
0.5 & 0.5 & 0 & 0\\
\end{pmatrix}
$$
Step 5:

node 2 will infect

Step 6:

lets choose node 3

Step 7: 
$$
A \cdot \mathbf{s}^T = 
\begin{pmatrix}
1 & 1 & 1 & 1\\
1 & 1 & 1 & 1\\
1 & 1 & 1 & 1\\
1 & 1 & 1 & 1\\
\end{pmatrix}
\cdot
\begin{pmatrix}
0\\
0\\
0\\
1\\
\end{pmatrix}
=
\begin{pmatrix}
1\\ 
1\\
1\\
1\\
\end{pmatrix}
$$

$$
(A \cdot \mathbf{s}^T)^T \circ \mathbf{c} = 
\begin{pmatrix}
1 & 1 & 1 & 1\\
\end{pmatrix}
\circ
\begin{pmatrix}
1 & 1 & 1 & 0\\
\end{pmatrix}
=
\begin{pmatrix}
1 & 1 & 1 & 0\\
\end{pmatrix}
$$

$$
T = \sum ((A \cdot \mathbf{s}^T)^T \circ \mathbf{c}) = 3
$$

Step 8:
$$
\Delta t = 1/3
$$

Iteration 3

Step 1:
$$
A \cdot \mathbf{s}^T =  
\begin{pmatrix}
1 & 1 & 1 & 1\\
1 & 1 & 1 & 1\\
1 & 1 & 1 & 1\\
1 & 1 & 1 & 1\\
\end{pmatrix}
\cdot
\begin{pmatrix}
0\\
0\\
0\\
1\\
\end{pmatrix}
=
\begin{pmatrix}
1\\
1\\
1\\
1\\
\end{pmatrix}
$$

Step 2:
$$
(A \cdot \mathbf{s}^T)^T \circ \mathbf{c} = 
\begin{pmatrix}
1 & 1 & 1 & 1\\
\end{pmatrix}
\circ
\begin{pmatrix}
1 & 1 & 1 & 0\\
\end{pmatrix}
=
\centering
\begin{pmatrix}
1 & 1 & 1 & 0\\
\end{pmatrix}
$$

Step 3:
$$
T = \sum ((A \cdot \mathbf{s}^T)^T \circ \mathbf{c}) = 3
$$

Step 4:
$$
\begin{pmatrix}
1 & 1 & 1 & 0\\
\end{pmatrix}
/ 3 = 
\begin{pmatrix}
1/3 & 1/3 & 1/3 & 0\\
\end{pmatrix}
$$

Step 5:

node 3 will infect

Step 6:

lets choose node 4

All are infected. And of algorithm.


\section{Program output}

Iteration 0

A = 
 [[0 1 1 1]
 [1 0 1 1]
 [1 1 0 1]
 [1 1 1 0]] 

s = 
 [[1]
 [0]
 [1]
 [1]] 

c =  [0 1 0 0] 

A x s o c =  [0 3 0 0] 

T =  3 

dt =  0.63135348587 

current time:  0

-------------------------

Iteration  1 

A = 
 [[0 1 1 1]
 [1 0 1 1]
 [1 1 0 1]
 [1 1 1 0]] 

s = 
 [[1]
 [0]
 [1]
 [1]] 

c =  [0 1 0 0] 

A x s o c =  [0 3 0 0] 

T =  3 

Probabilities to infect =  [0. 1. 0. 0.] 

node  1  will infect

Probabilities to get infected =  [0.33333333 0.         0.33333333 0.33333333] 

node  3  gets infected

A = 
 [[0 1 1 1]
 [1 0 1 1]
 [1 1 0 1]
 [1 1 1 0]] 

s = 
 [[1]
 [0]
 [1]
 [0]] 

c =  [0 1 0 1] 

A x s o c =  [0 2 0 2] 

T =  4 

dt =  0.484454435736 

current time :  0.63135348587

-------------------------

Iteration  2 

A = 
 [[0 1 1 1]
 [1 0 1 1]
 [1 1 0 1]
 [1 1 1 0]] 

s = 
 [[1]
 [0]
 [1]
 [0]] 

c =  [0 1 0 1] 

A x s o c =  [0 2 0 2] 

T =  4 

Probabilities to infect =  [0.  0.5 0.  0.5] 

node  3  will infect

Probabilities to get infected =  [0.5 0.  0.5 0. ] 

node  2  gets infected

A = 
 [[0 1 1 1]
 [1 0 1 1]
 [1 1 0 1]
 [1 1 1 0]] 

s = 
 [[1]
 [0]
 [0]
 [0]] 

c =  [0 1 1 1] 

A x s o c =  [0 1 1 1] 

T =  3 

dt =  0.317254365758 

current time :  1.115807921606

-------------------------

Iteration  3 

A = 
 [[0 1 1 1]
 [1 0 1 1]
 [1 1 0 1]
 [1 1 1 0]] 

s = 
 [[1]
 [0]
 [0]
 [0]] 

c =  [0 1 1 1] 

A x s o c =  [0 1 1 1] 

T =  3 

Probabilities to infect =  [0.         0.33333333 0.33333333 0.33333333] 

node  3  will infect

Probabilities to get infected =  [1. 0. 0. 0.] 

node  0  gets infected

A = 
 [[0 1 1 1]
 [1 0 1 1]
 [1 1 0 1]
 [1 1 1 0]] 

s = 
 [[0]
 [0]
 [0]
 [0]] 

c =  [1 1 1 1] 

A x s o c =  [0 0 0 0] 

T =  0 

dt =  inf 

current time :  1.433062287364
\end{appendices}

\end{document}